\documentclass[
 aip, 
 amsmath,amssymb,
 reprint, superscriptaddress,
 citeautoscript,  
]{revtex4-1}
\usepackage[utf8]{inputenc}
\usepackage{graphicx}
\usepackage{float}
\usepackage{enumitem}
\usepackage{bm}
\usepackage{bbold}
\usepackage{cases}
\usepackage{upgreek}
\usepackage[usenames, dvipsnames]{color}
\usepackage[colorlinks=true, citecolor=Plum, linkcolor=Mulberry, urlcolor=Orchid]{hyperref}
\usepackage{hyperref}
\usepackage{braket}
\usepackage[normalem]{ulem}

\usepackage{textgreek}
\usepackage{gensymb}
\usepackage{dcolumn}

\newcommand \mc[1] { \mathcal{#1} }
\newcommand \rmm[1]  { \textrm{#1} }

\makeatletter
\def\@email#1#2{%
 \endgroup
 \patchcmd{\titleblock@produce}
  {\frontmatter@RRAPformat}
  {\frontmatter@RRAPformat{\produce@RRAP{*#1\href{mailto:#2}{#2}}}\frontmatter@RRAPformat}
  {}{}
}%
\makeatother

\begin{document}

\preprint{AIP/123-QED}

\title{Space-local memory in generalized master equations: Reaching the thermodynamic limit for the cost of a small lattice simulation}

\author{Srijan Bhattacharyya}
\affiliation{Department of Chemistry, University of Colorado Boulder, Boulder, CO 80309, USA}

\author{Thomas Sayer}
\affiliation{Department of Chemistry, University of Colorado Boulder, Boulder, CO 80309, USA}
\affiliation{Department of Chemistry, Durham University, Durham DH1 3LE, United Kingdom}

\author{Andr\'{e}s Montoya-Castillo}
\homepage{Andres.MontoyaCastillo@colorado.edu}
\affiliation{Department of Chemistry, University of Colorado Boulder, Boulder, CO 80309, USA} 

\date{\today}

\begin{abstract}
The exact quantum dynamics of lattice models can be computationally intensive, especially when aiming for large system sizes and extended simulation times necessary to converge transport coefficients. By leveraging finite memory times to access long-time dynamics using only short-time data, generalized master equations (GMEs) can offer a route to simulating the dynamics of lattice problems efficiently. However, such simulations are limited to small lattices whose dynamics exhibit finite-size artifacts that contaminate transport coefficient predictions. To address this problem, we introduce a novel approach that exploits finite memory in time \textit{and} space to efficiently predict the many-body dynamics of dissipative lattice problems involving short-range interactions. This advance enables one to leverage the short-time dynamics of small lattices simulate arbitrarily large systems over long times. We demonstrate the strengths of this method by focusing on nonequilibrium polaron relaxation and transport in the dispersive Holstein model, successfully simulating lattice dynamics in one and two dimensions free from finite-size effects, reducing the computational expense of such simulations by multiple orders of magnitude. Our method is broadly applicable and provides an accurate and efficient means to investigate nonequilibrium relaxation with microscopic resolution over mesoscopic length and time scales that are relevant to experiment.
\end{abstract}

\maketitle

\section{Introduction}
\vspace{-4pt}

Lattice models play a key role in understanding physical and chemical phenomena. For instance, the Holstein~\cite{holstein_studies_1959_part1, holstein_studies_1959} and Fr\"ohlich~\cite{frohlich_electrons_1954} models shed light on polaron formation and electrical transport in semiconductors~\cite{hulea_tunable_2006, fetherolf_unification_2020}, the Hubbard model~\cite{hubbard_electron_1963} helps elucidate the mechanisms of high-temperature superconductivity~\cite{arrigoni_mechanism_2004}, and the Ising model~\cite{ising_beitrag_1925} is used to interrogate magnetism~\cite{newell_theory_1953} and phase transitions~\cite{dziarmaga_dynamics_2005}. However, while modern algorithms can efficiently simulate the quantum dynamics of small lattices over short times~\cite{tanimura_time_1989, makri_tensor_1995, wang_systematic_2001, thoss_self-consistent_2001, suess_hierarchy_2014, wang_multilayer_2015, de_vega_thermofield-based_2015, strathearn_efficient_2018, tamascelli_efficient_2019, schroder_tensor_2019, xie_time-dependent_2019, kloss_multiset_2019, makri_small_2021,  bose_multisite_2022, gribben_exact_2022, fux_tensor_2023, lacroix_mpsdynamicsjl_2024}, reaching sufficiently large systems and long times to compare to experiments remains a fundamental challenge. is is because these methods often scale exponentially or, at best, polynomially with lattice size and simulation time, rendering the thermodynamic limit inaccessible. 

The severity of this limitation becomes clear when calculating dynamic properties, e.g., conductivities, viscosities, and diffusion constants, which are sensitive to finite-size effects~\cite{kikugawa_hydrodynamic_2015, simonnin_diffusion_2017, cox_interfacial_2018, jamali_finite-size_2018, samanta_finite_2018, bertini_finite-temperature_2021,celebi_finite-size_2021, cox_dielectric_2022}. For example, finite-size effects can cause simulations to underestimate diffusion constants of polymers near the glass-transition~\cite{ray_finite-size_1994}, relaxation times of glass-forming liquids~\cite{berthier_finite-size_2012}, the Curie temperature for Ni nanoparticles~\cite{dos_santos_size-dependent_2024}, and the diffusion constant and viscosity of model fluids~\cite{yeh_system-size_2004}; overestimate the critical fermion-phonon coupling causing the metal-to-Peierls phase transition in a Holstein-Hubbard lattice~\cite{hebert_one-dimensional_2019}; and yield apparently non-converging mobilities of dispersive Holstein polarons~\cite{bhattacharyya_anomalous_2024}. These examples reveal the need of computing the dynamics of lattice models in thermodynamically large systems over long timescales.

Generalized Master Equations (GMEs) have emerged as a powerful tool for reducing the computational cost of dynamical simulations~\cite{shi_new_2003, shi_semiclassical_2004, zhang_nonequilibrium_2006, cohen_memory_2011, cohen_generalized_2013, kelly_efficient_2013, kidon_exact_2015, kelly_accurate_2015, kelly_generalized_2016, montoya-castillo_approximate_2016, montoya-castillo_approximate_2016, pfalzgraff_efficient_2019, mulvihill_combining_2019, ng_nonuniqueness_2021, mulvihill_simulating_2022, amati_quasiclassical_2022, lyu_tensor-train_2023, wang_simulating_2023, sayer_compact_2023, sayer_efficient_2024, sayer_generalized_2024}. GMEs are exact non-Markovian equations of motion for nonequilibrium averages, correlation functions, and even multi-time correlators of select variables that encapsulate the effects of an environment into a memory kernel~\cite{nakajima_quantum_1958, zwanzig_ensemble_1960, mori_transport_1965}. In dissipative systems, the memory kernel decays to zero over a finite memory lifetime, which can be shorter than the relaxation time of the desired correlation function. Thus, in principle, one can use a reference simulation over the memory lifetime to construct a GME that predicts the dynamics of the desired correlation function to arbitrarily long times. This temporal truncation of memory at its lifetime can reduce the computational cost of simulating the quantum or classical dynamics of complex many-body systems in different problems, including charge transfer reactions in solution~\cite{pfalzgraff_nonadiabatic_2015, liu_combining_2024}, protein folding~\cite{cao_advantages_2020, dominic_building_2023, cao_integrative_2023}, nonlinear spectroscopy~\cite{ivanov_extension_2015, sayer_efficient_2024}, and transport~\cite{yan_theoretical_2019, bhattacharyya_mori_2024}. However, to construct a GME from a short-time reference simulation, it must satisfy two conditions: (1) the simulation time must span the memory kernel lifetime, and (2) the reference calculation must be performed in the same system whose dynamics one intends to interrogate with the GME. If one constructs a GME using a small lattice simulation, particles encounter the lattice boundaries and manifest finite-size effects: one reduces the cost but still obtains the wrong answer. Hence, one must be able to afford an admittedly short-time reference simulation, but of a thermodynamically large lattice. The poor scaling of dynamical methods with system size renders this calculation at best impractical and at worst impossible.

Here, we propose a novel approach to lattice problems that exploits our observation that certain GME formulations display a \textit{finite spatial memory} to motivate truncating memory in time \textit{and} space. 
This allows us to employ short-time reference simulations of small lattices to generate the exact quantum dynamics of thermodynamically large lattices over arbitrarily long times for the cost of only the small reference calculation. We demonstrate the strengths of this method by applying it to nonequilibrium polaron formation and transport in dispersive Holstein lattices. Enabled by our space-local GME, we simulate, for the first time, the exact nonequilibrium quantum dynamics of small polaron formation, relaxation, and flow in thermodynamically large one-(1D) and two-dimensional (2D) lattices with up to $900$ sites over $100$ ps, free of finite-size contamination. Our method is model-agnostic and can be expected to enable the efficient investigation of nonequilibrium excitation dynamics in dissipative lattices displaying local interactions.

\section{Small polaron lattice}
\vspace{-4pt}

The dispersive Holstein model offers a physically transparent description of small polaron formation and transport in many semiconductors displaying strong carrier-phonon interactions, including organic crystals~\cite{cheng_unified_2008}, polymers~\cite{qarai_understanding_2021}, and nanomaterials~\cite{mousavi_effects_2012}. This model describes the migration of charge carriers (electrons or holes) or excitons that interact with their environment, influencing local lattice (nuclear) motions. This interaction causes the material to deform around the carrier, forming a \textit{polaron}. 

In the dispersive Holstein model,
\begin{equation}
    \hat{H} = \hat{H}_s + \hat{H}_{ph} + \hat{H}_{s-ph}, \label{eq:full-ham}
\end{equation}
where $\hat{H}_s$ describes the electronic carriers: 
\begin{equation}
    \hat{H}_s = \sum_{i}^{N} \epsilon_i  \hat{a}_i^\dag \hat{a}_i + \sum_{\langle ij \rangle}^{N} v_{ij} \hat{a}_i^\dag \hat{a}_j. \label{eq:free-sys-ham}
\end{equation}
Here $\epsilon_i$ represents the on-site energy, $v_{ij}$ the hopping integral connecting the $i^\rmm{th}$ and $j^\rmm{th}$ sites,  $N$ the number of sites on the lattice, and $\langle ij \rangle$ restricts the sum to nearest neighbor coupling. While a carrier couples to a single local phonon mode per lattice site in the classic Holstein model~\cite{holstein_studies_1959_part1, holstein_studies_1959, mahan_many-particle_2000}, we opt for the dispersive version where a carrier couples to a continuum of local phonon modes, which more faithfully represents condensed phase systems~\cite{mayers_how_2018, nematiaram_modeling_2020, krylow_ultrafast_2020, li_excitation-wavelength-dependent_2023} and has a full harmonic phonon environment per lattice site,
\begin{equation}
    \hat{H}_{ph} = \frac{1}{2} \sum_{i} \sum_{\alpha} \big[ \hat{P}_{i,\alpha}^2 +  \omega_{i\alpha}^2 \hat{X}_{i,\alpha}^2 \big]. \label{eq:free-bath-ham}
\end{equation}
$\hat{X}_{i,\alpha}$ and $\hat{P}_{i,\alpha}$ are mass-weighted positions and momenta for the $\alpha^\rmm{th}$ phonon connected to the carrier on site $i$. The carrier-phonon coupling is linear in the phonon coordinates,
\begin{equation}
    \hat{H}_{s-ph} = \sum_i \sum_{\alpha} c_{i, \alpha} \hat{X}_{i,\alpha}  \hat{a}_i^\dag \hat{a}_i.
\end{equation}
This carrier-phonon coupling is fully characterized by the spectral density  $J_i(\omega) = \frac{\pi }{2} \sum_{\alpha} \frac{c_{i,\alpha}^2} {\omega_{i\alpha}} \delta(\omega -\omega_{i\alpha}).$

Consistent with previous works~\cite{yan_theoretical_2019, fetherolf_unification_2020, bhattacharyya_anomalous_2024}, we focus on the dilute limit (one-carrier manifold) of homogeneous lattices, making all our parameters site-independent. Without loss of generality, we set  $\epsilon_i = 0$, nearest-neighbor hopping integral $v_{ij} = v$, intersite distances $r_0 = 5$\,\AA, and all spectral densities to an Ohmic-Debye form commonly used to mimic dissipation in the condensed phase~\cite{weiss_quantum_2012}: $J(\omega) = \eta \omega_c \omega/(\omega^2 + \omega_c^2).$ Here, $\eta/2$ is the lattice reorganization energy of an occupied lattice site, and $1/\omega_c$ the phonon environment's decorrelation time. 

In our simulations, we track the population matrix $\boldsymbol{C}(t)$, where $ C_{i,j}(t) = \mathrm{Tr} [\hat{a}_i^\dag \hat{a}_i e^{-\rmm{i}\mathcal{L}t} \hat{\rho}_j(0)]$ with initial condition $\hat{\rho_j}(0) = \hat{a}_j^{\dagger}\hat{a}_j e^{-\beta \hat{H}_{ph}}/\mathrm{Tr}[e^{-\beta \hat{H}_{ph}}]$, which corresponds to creating an excitation at site $ j $ at $ t = 0 $ when all the phonon modes are in thermal equilibrium with the electronic ground state at inverse temperature $\beta = 1/ k_{\mathrm{B}} T$. $ C_{i,j}(t)$ quantifies the probability of finding the polaron at site $ i $ at time $t$ given that it was initiated at site $ j $. In homogeneous systems with equivalent sites and periodic boundary conditions, $ C(t) $ exhibits translational invariance, i.e., $ C_{i,i+l\mod N }(t) = C_{j,j+l \mod N}(t)$ for all $i,j,l$. This allows one to identify all distinct elements of this matrix by their relative site index, $|i-j| = k$, allowing one to rewrite $ C_{i,j}(t) = C_{|i-j|}(t) = C_k(t) $. Hence, one only needs to perform a single simulation starting the carrier at any particular site, $j$, and record the time-dependent probability of finding the carrier at all lattice sites, $i$, as a function of time to construct $\boldsymbol{C}(t)$.

\begin{figure}[t]
\begin{center} 
\vspace{-3pt}
    \resizebox{.44\textwidth}{!}{\includegraphics[trim={0pt 0pt 0pt 0pt},clip]{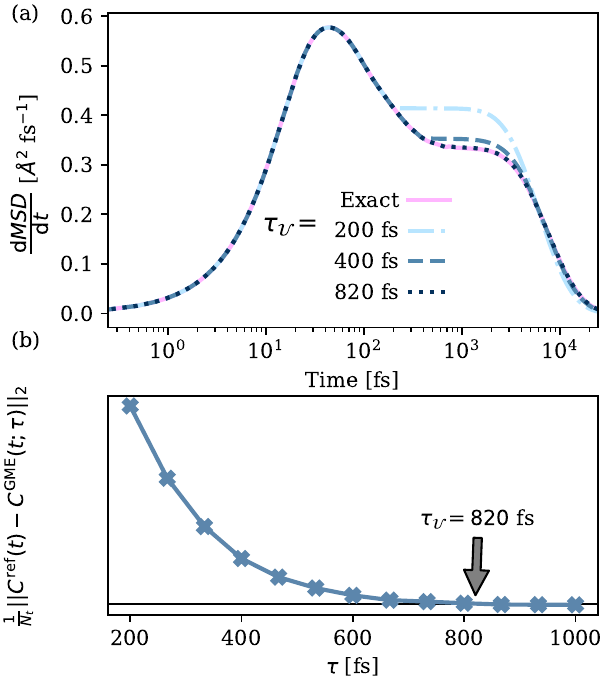}}
    \vspace{-20pt}
\end{center}
\caption{\label{fig:time-local} Nonequilibrium polaron dynamics on a $20$-site 1D lattice with $\eta=323$~cm$^{-1}$, $v = 50$~cm$^{-1}$ and $\omega_c= 41$~cm$^{-1}$. (a) Comparison of $\mathrm{d}MSD/\mathrm{d}t$ between the exact (pink) and TL-GME dynamics generated with different lifetimes (broken blue lines). (b) $||\mathrm{L}||_2$ error for the TL-GME prediction as a function of the proposed time cutoff reveals $\tau_{\mc{U}} \approx 820$~fs. The error is normalized by the number of time points predicted by the GME, $N_t$. }
\vspace{-5pt}
\end{figure}

\begin{figure*}
\begin{center} 
\vspace{-3pt}
    \resizebox{.9\textwidth}{!}{\includegraphics[trim={0pt 0pt 0pt 0pt},clip]{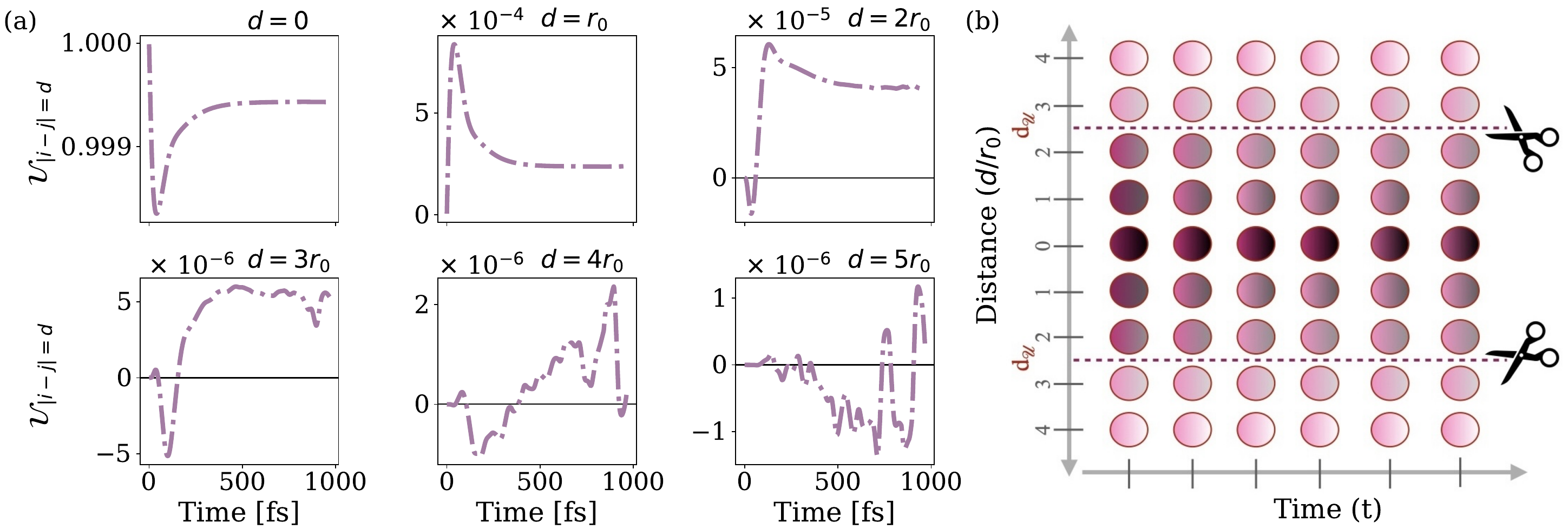}}
    \vspace{-3pt}
\end{center}
\caption{\label{fig:schematic}(a) Elements of $\boldsymbol{\mathcal{U}} (t) $ as a function of lattice distance $d$ for the dispersive Holstein lattice with the same parameters as Fig.~\ref{fig:time-local}. As inter-site distance, $d$, increases, the elements of $\boldsymbol{\mathcal{U}} (t)$ become smaller. (b) Schematic representation of the spatial locality in the generator, $\boldsymbol{\mathcal{U}} (t) $, where elements beyond a characteristic distance $\mathrm{d}_{\mathcal{U}}$ become negligible. The color intensity quantifies the magnitude of the elements and the scissors indicate the spatial truncation of the generator.}
\vspace{-5pt}
\end{figure*}

To characterize transport, we compute the polaron's mean squared displacement, $MSD(t) = r_0^2 \sum_k  k^2 C_k(t)$, which determines its diffusion constant $D$ via the time derivative when the $MSD$ scales linearly in time:
\begin{equation}\label{eq:def-d}
    D = \frac{1}{2N_d} \lim_{t \to \infty} \frac{\mathrm{d}MSD}{\mathrm{d}t}.
\end{equation}
Here $ N_d $ indicates the lattice dimension. We calculate the reference $\boldsymbol{C}(t)$ using the numerically exact Hierarchical Equation of Motion (HEOM)~\cite{shi_efficient_2009, liu_reduced_2014, song_time_2015} under periodic boundary conditions at $T = 300$ K for all simulations in this work (see App.~\ref{HEOM-details} for additional details).

\section{Method Development \& Analysis}
\vspace{-4pt}

We begin with the $\frac{\mathrm{d}MSD}{\mathrm{d}t}$ dynamics of dispersive Holstein polarons on a 1D lattice. To align with previous work on organic semiconductors~\cite{song_new_2015, bhattacharyya_anomalous_2024}, we take $\eta = 323$~cm$^{-1}$, $v = 50$~cm$^{-1}$ and  $\omega_c= 41$~cm$^{-1}$. Beyond encoding the diffusion constant, $\frac{\mathrm{d}MSD}{\mathrm{d}t}$ provides insights into polaron formation, the transition from nonequilibrium relaxation to diffusive transport, and the onset of finite-size contamination~\cite{bhattacharyya_anomalous_2024}. For example, in the reference (HEOM) dynamics for the 20-site lattice (solid pink line) in Fig.~\ref{fig:time-local} (a), the initial hump at $\sim 100$~fs reports on far-from-equilibrium lattice reorganization from polaron formation, and is followed by a plateau that suggests diffusive motion between $800$ and $2000$~fs. While this region is not truly flat, it approximates the diffusion constant~\cite{bhattacharyya_anomalous_2024}. This putative plateau falls off at $\sim 2000$ fs, marking the onset of finite-size effects. We denote this onset timescale as $\tau_R$. 

We significantly reduce the computational cost of this simulation by adopting a GME for the population matrix. Specifically, we adopt the integrated time-local (TL) master equation~\cite{sayer_compact_2023} (see App.~\ref{app:pop-GME}):
\begin{equation}\label{TL-GME}
    \boldsymbol{C}(t + \delta t) = \boldsymbol{\mathcal{U}}(t) \boldsymbol{C}(t),
\end{equation}
where $ \boldsymbol{\mathcal{U}}(t) $ is the generator of the non-Markovian dynamics for $\boldsymbol{C}(t)$. In dissipative lattice systems, $\boldsymbol{\mathcal{U}} $ can be expected to have a finite memory lifetime $\tau_{\mathcal{U}}$, after which it becomes a time-independent matrix $\boldsymbol{\mathcal{U}}(\tau_{\mathcal{U}})$, indicating the onset of Markovianity. At this point, Eq.~\ref{TL-GME} simplifies to a time-independent rate equation and one can efficiently generate long-time dynamics through matrix multiplication:
\begin{equation}\label{TL-GME2}
    \boldsymbol{C}(\tau_{\mathcal{U}} + n\delta t) = [\boldsymbol{\mathcal{U}}(\tau_{\mathcal{U}})]^n \boldsymbol{C}(\tau_{\mathcal{U}}),
\end{equation}
where $n \in \mathbb{N}$. We use the reference HEOM dynamics to calculate the generator using 
\begin{equation}\label{eq-generator}
    \boldsymbol{\mathcal{U}}(t) = \boldsymbol{C}(t+\delta t) [\boldsymbol{C}(t)]^{-1}.
\end{equation}
To identify the generator lifetime~\cite{sayer_compact_2023}, we truncate $\boldsymbol{\mathcal{U}}(t)$ at sample lifetimes $\tau$ and use the resulting generator to produce TL-GME dynamics, from which we compute the $||\mathrm{L}||_2$ norm of the difference between the reference dynamics and the TL-GME dynamics. We find that $\tau_{\mathcal{U}} = 820$~fs, which is when the error metric in Fig~\ref{fig:time-local} (b) plateaus near zero, allowing the resulting TL-GME dynamics (dotted blue line) to agree with the reference dynamics (pink line) in Fig~\ref{fig:time-local} (a). Employing a smaller lifetime for the generator causes the TL-GME to predict inaccurate dynamics, as illustrated by the dashed blue lines in Fig.~\ref{fig:time-local} (a). Hence, one only needs a reference simulation up to the generator lifetime $\tau_{\mc{U}} = 820$~fs to generate the dynamics over $25$~ps---more than an order of magnitude longer---via simple matrix multiplication. Yet, while combining exact simulations with the TL-GME can reduce the computational cost of generating long-time dynamics in such lattice problems, the finite size of the lattice invariably poisons the dynamics once the polaron reaches the boundary. This prompts us to ask: can one reach large system sizes without incurring the cost of reference simulations over increasingly larger lattices that become prohibitively expensive? 

\textbf{\textit{Spatial memory:}} We address this fundamental problem by leveraging the concept of \textit{spatial memory} to develop a \textit{space-local (SL)} GME that uses only a short-time simulation of a small lattice to predict the dynamics of thermodynamically large lattices over arbitrary times. We begin by noting that, like $\boldsymbol{C}(t)$, $\boldsymbol{\mathcal{U}}(t)$ exhibits translational invariance in homogeneous lattices, allowing us to use relative indices to identify the elements. Figure~\ref{fig:schematic}~(a) shows that \textit{as intersite distance increases, the elements of $\boldsymbol{\mathcal{U}}(t)$ become progressively smaller}, $ \mathcal{U}_{d+2} < \mathcal{U}_{d+1} < \mathcal{U}_{d} $, suggesting that our TL-GME generator exhibits decaying spatial memory. Further, one may posit that generator elements connecting sites separated by a distance greater than a \textit{characteristic memory distance}, $\mathrm{d}_{\mc{U}}$, become negligibly small, allowing one to set them to zero. We illustrate this concept of spatial memory truncation in Fig.~\ref{fig:schematic}~(b), where we discard the elements of $\boldsymbol{\mathcal{U}}(t)$ beyond a particular distance, $d = \mathrm{d}_{\mc{U}}$, as they become negligibly small. 

\begin{figure}[b]
\begin{center} 
\vspace{-3pt}
    \resizebox{.44\textwidth}{!}{\includegraphics[trim={0pt 0pt 0pt 0pt},clip]{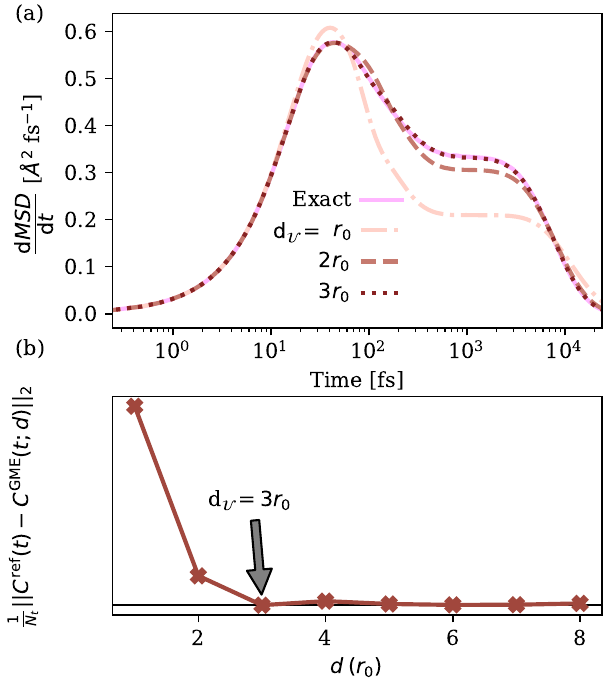}}
    \vspace{-15pt}
\end{center}
\caption{\label{fig:space-local} Demonstration of the SL-GME's ability to capture the polaron dynamics on the $20$-site 1D lattice with the same parameters as Fig.~\ref{fig:time-local}. (a) Comparison of $\mathrm{d}MSD/\mathrm{d}t$ dynamics obtained with exact HEOM (pink solid line) and SL-GME (brown broken lines) dynamics. (b) Deviation (error) of the SL-GME dynamics from the exact dynamics as a function of distance cutoff, revealing that $\mathrm{d}_{\mc{U}} = 3r_0$. Error is defined in the same way as of Fig.~\ref{fig:time-local} (b) }
\vspace{-5pt}
\end{figure}

We turn to testing the validity of SL-GME. We begin by examining the $||\mathrm{L}||_2$ error metric for the SL-GME as a function of the proposed memory distance cutoff, $d$, to identify the characteristic memory distance, $\mathrm{d}_{\mathcal{U}}$. 
Figure~\ref{fig:space-local}~(a) shows that the SL-GME reproduces the reference dynamics (solid pink line) when $\mathrm{d}_{\mathcal{U}} = 3 r_0$ (dotted brown line), but not for smaller cutoff distances. The plateau in the error as a function of distance cutoff in Fig.~\ref{fig:space-local} (b) confirms that $\mathrm{d}_{\mathcal{U}} = 3 r_0$, achieving an average error of less than $1\%$. These results thus support the proposal of a memory distance cutoff that preserves accuracy in the SL-GME dynamics. 

Since obtaining the generator up to its lifetime $ \tau_{\mathcal{U}} $ allows one to simulate the dynamics for \textit{all time}, one may hypothesize that constructing the generator up to its characteristic memory distance $\mathrm{d}_{\mathcal{U}} $ should enable one to simulate a lattice \textit{of any size}, albeit of the same dimension. This is the central insight of our \textit{space- and time-local GME (STL-GME)}. To achieve this in a 1D lattice, we propose \textit{augmenting} the dimension of $\boldsymbol{\mathcal{U}}(t)$, which is a 3-tensor with dimensions $[N, N, tsteps]$, to $[M, M, tsteps]$, where $M> N$ and all new elements are assigned a value of $0$. Further truncating the temporal dimension of our augmented generator at time $ \tau_{\mathcal{U}} $ enables us to easily generate the dynamics after $ \tau_{\mathcal{U}} $ via simple matrix multiplication. This spatial and temporal truncation of $\boldsymbol{\mathcal{U}}(t)$ followed by its spatial augmentation constitutes our STL-GME. We offer details of its implementation in App.~\ref{app:stl-gme}. The resulting STL-GME should thus offer a route to employ the dynamics of a small lattice over short times to simulate the behavior of a thermodynamically large lattice across arbitrarily long timescales. 

\begin{figure}[b]
\begin{center} 
\vspace{-3pt}
    \resizebox{.5\textwidth}{!}{\includegraphics[trim={0pt 0pt 0pt 0pt},clip]{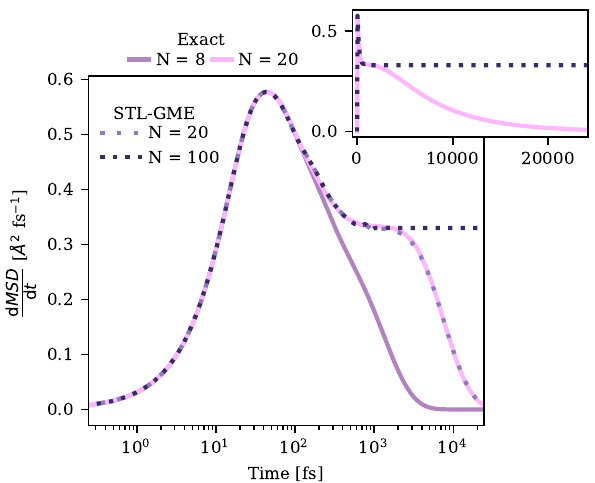}}
    \vspace{-30pt}
\end{center}
\caption{\label{fig:space-time-local} Benchmarking of our STL-GME for dispersive Holstein dynamics for a 20-site (pink line) 1D lattice with $\eta=323$~cm$^{-1}$, $v = 50$~cm$^{-1}$, $\omega_c= 41$~cm$^{-1}$. We construct our STL-GME generator using a reference 8-site (purple line) simulation, which we then augment in size to simulate GME dynamics for 20- and 100-site lattices (broken blue lines). Inset: normal-scale plot of $\mathrm{d}MSD/\mathrm{d}t$ for the 20- and 100-site lattices. The 20-site dynamics exhibit finite-size effects around $\sim 2000$~fs, while the 100-site lattice does not manifest them over the simulated times.}
\vspace{-5pt}
\end{figure}

We test our STL-GME's performance by turning again to the Holstein lattice in Figs.~\ref{fig:time-local} and~\ref{fig:space-local}. Since $\boldsymbol{\mathcal{U}}$ is an \textit{intrinsic} property of the dynamics, one can extract it from short-time $t \leq \tau_{\mc{U}}$ reference dynamics of a lattice with $N \geq 2\mathrm{d}_{\mc{U}}+1$ sites. One can then employ $\boldsymbol{\mathcal{U}}(t)$ to predict relaxation dynamics over \textit{arbitrarily large spaces and times, removing all finite-size artifacts}. We validate this idea in Fig.~\ref{fig:space-time-local}, where we construct the generator for our STL-GME dynamics for a lattice with $N=20$ sites using a reference simulation of a lattice with only $N=8>2\times (\mathrm{d}_\mc{U}=3)+1$ sites (solid purple line). Our size-augmented STL-GME in Fig.~\ref{fig:space-time-local} (light purple dots) reproduces the reference dynamics for a lattice with $N = 20$ sites (solid pink line). In fact, the STL-GME can access the dynamics of arbitrarily large systems, indefinitely delaying the onset of finite-size effects. As we show in the inset of Fig.~\ref{fig:space-time-local}, a $20$-site lattice exhibits finite-size artifacts around $ \sim 2000 $~fs, whereas our STL-GME calculation for the $100$-site system shows no signs of finite-size effects, even by 25~ps.

Yet, that $N$ can be so small in the reference calculation used to construct the generator in Fig.~\ref{fig:space-time-local} is surprising. Here, $\tau_\mc{U}\simeq 800$~fs, but for an 8-site lattice, finite-size effects manifest in $\mathrm{d}MSD/\mathrm{d}t$ at $\sim 300$~fs, i.e., $\tau_R \simeq 300$ fs. However, despite $\tau_\mc{U} > \tau_R$, the STL-GME and reference calculations agree, meaning the STL-GME correctly simulates the larger lattice and is not contaminated by the smallness of the reference system. 

How can one understand this seeming contradiction where the generator lifetime, $\tau_\mc{U}$, can be longer than $\tau_R$, the onset of finite-size effects in $\mathrm{d}MSD/\mathrm{d}t$? Immediately at $\tau_R$, the finite-size effect contaminates only the elements of $\boldsymbol{\mc{U}}$ connecting the origin to the most distant sites. This is true because the population density is conserved and its current local. In time, this artifact spreads to elements of $\boldsymbol{\mc{U}}$ progressively closer to the origin, as the density interferes with itself. One might hypothesize that finite-size effects only manifest in the STL-GME when the elements of $\boldsymbol{\mc{U}}$ that survive the spatial truncation have been contaminated---a time longer than $\tau_R$. 

\begin{figure}[t]
\begin{center} 
\vspace{-3pt}
    \resizebox{.49\textwidth}{!}{\includegraphics[trim={0pt 0pt 0pt 0pt},clip]{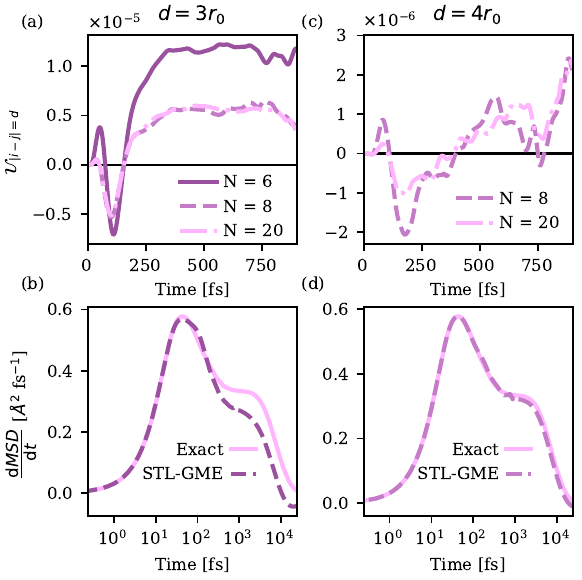}}
    \vspace{-25pt}
\end{center}
\caption{\label{fig:u-diff} Convergence of generator elements as a function of lattice size and characteristic memory distance, $\mathrm{d}_{\mc{U}}$, for dispersive Holstein parameters $v = 50$~cm$^{-1}$, $\eta = 323$~cm$^{-1}$, $\omega_c = 41$~cm$^{-1}$. (a) $\mc{U}(d = 3r_0,t)$ as a function of lattice size. (b) Impact of adopting a generator from a too-small lattice, $\mc{U}_{N=6}(d=3 r_0, t)$. (c) $\mc{U}(d=4r_0,t)$ as a function of lattice size. (d) Impact of adopting a (contaminated) generator from a too-large choice of cutoff distance for the lattice size, $\mc{U}_{N=8}(d=4r_0, t)$.}
\vspace{-5pt}
\end{figure}

To confirm this hypothesis, in Fig.~\ref{fig:u-diff} we compare the STL-GME generator and dynamics constructed from reference data for two different lattice sizes with varying cutoff distances. The parameter regime is the same as in Fig.~\ref{fig:space-time-local}, where we found $\mathrm{d}_{\mc{U}}=3r_0$ and $\tau_{\mc{R}} = 300$~fs, thus requiring a minimal lattice of $N=8$. Keeping $\tau_{\mc{U}} = 800$~fs constant, we consider two complementary cases. First, for a too-small 6-site lattice, we know that obtaining a result free of finite-size effects is impossible because the periodic boundary is always closer than $3$ sites from the initial polaron position. Indeed, $\mc{U}_{N=6}(d=3r_0, t)$ differs significantly from the benchmark $\mc{U}_{N=20}(d=3r_0, t)$ (Fig.~\ref{fig:u-diff} (a)), with the former producing STL-GME dynamics that deviate from the reference simulation for a 20-site lattice (Fig.~\ref{fig:u-diff} (b)). Second, we consider the previously sufficient 8-site lattice, but increase the cutoff to $d_{\mc{U}}=4r_0$. Since $N<2d_\mc{U}+1$, we predict finite-size contamination in $\mc{U}_{N=8}(d=4r_0, t)$. This is evident when comparing with the benchmark $\mc{U}_{N=20}(d=4r_0, t)$ (Fig.~\ref{fig:u-diff} (c)) and from the STL-GME dynamics, which again deviate from the reference (Fig.~\ref{fig:u-diff} (d)). That is, while the 8-site lattice generator's first 4 entries are correct for this $\tau_\mc{U}$, the fifth element is contaminated by finite-size artifacts, degrading the predicted dynamics. In both of these pathological cases, truncation in time and space happens \textit{after} knowledge of the lattice's finite size reaches those elements in $\boldsymbol{\mc{U}}$ that we keep. Therefore, a lattice with $N=8$ sites converges to the right value when one truncates spatially at $d_{\mc{U}}=3r_0$, as we confirm by comparing $\mc{U}_{N=8}(d=3r_0, t)$ and $\mc{U}_{N=20}(d=3r_0, t)$, and the resulting STL-GME prediction with the reference results for a 20-site lattice, shown in Fig.~\ref{fig:space-time-local}. Hence, the STL-GME is even more efficient in accessing the dynamics of thermodynamically large systems from small reference calculations than one might have expected based on simple physical arguments.

\textit{\textbf{Applicability:}} Thus far, we have assumed only nearest-neighbor coupling, raising questions about the extent to which our conclusions depend on the locality of the Hamiltonian. For example, if the electronic coupling extends further, over second and third nearest neighbors, do the cutoffs extend by $2r_0$, or is there a more pernicious, qualitative change? 

\begin{figure}[t]
\begin{center} 
\vspace{-3pt}
    \resizebox{.49\textwidth}{!}{\includegraphics[trim={0pt 0pt 0pt 0pt},clip]{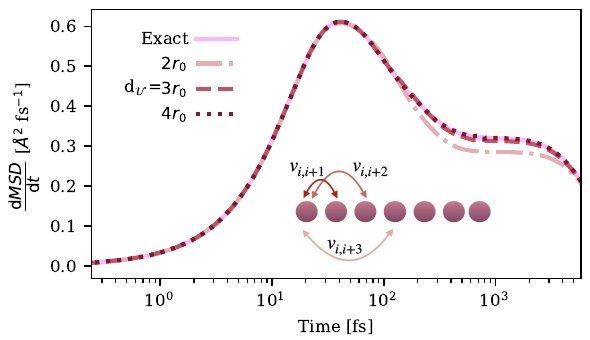}}
    \vspace{-28pt}
\end{center}
\caption{\label{fig:more-nn} Applicability of the STL-GME for a dispersive Holstein lattice with short-range, non-local couplings. Hamiltonian parameters: $\eta=323$~cm$^{-1}$, $v (r_0) = 50$~cm$^{-1}$,  $v (2 r_0) \approx 6.8$~cm$^{-1}$, $v (3r_0) \approx 1.0$~cm$^{-1}$, $\omega_c= 41$~cm$^{-1}$.}
\vspace{-10pt}
\end{figure}

To probe this potential sensitivity to non-locality, we recalculate our dispersive Holstein dynamics with beyond-nearest-neighbor electronic couplings for the parameter regime of Fig.~\ref{fig:space-local}, where we had identified $\mathrm{d}_{\mc{U}} = 3r_0$ when considering only nearest-neighbor coupling. Motivated by atomic orbital decays, we consider interactions that decrease exponentially with inter-site distance up to third nearest neighbors, as shown in Fig.~\ref{fig:more-nn}, with $v(d) = v e^{-a(d- r_0)}$, $a = 2\mathrm{r_0}^{-1}$. Even in this non-locally interacting problem, Fig.~\ref{fig:more-nn} shows that our STL-GME captures the reference dynamics with $\mathrm{d}_\mc{U} = 4 r_0$. This change in cutoff distance of one lattice spacing is even less severe than the $2r_0$ one may have anticipated. Thus, our STL-GME efficiently handles Hamiltonians with non-local, albeit short-range, interactions. While more general models include long-range, through-space effects like carrier-carrier interaction, screening in the condensed phase~\cite{debye_theorie_1923, kirkwood_statistical_1936, rowlinson_yukawa_1989, ludwig_describing_2018, rotenberg_underscreening_2018, hartel_anomalous_2023, yang_solvent_2023, dinpajooh_detecting_2024} usually allows for a truncation that can be modeled similarly to the modification tested here and should therefore preserve the space-local arguments central to our STL-GME method.

\begin{figure}[b]
\begin{center} 
\vspace{-3pt}
    \resizebox{.4\textwidth}{!}{\includegraphics[trim={0pt 0pt 0pt 0pt},clip]{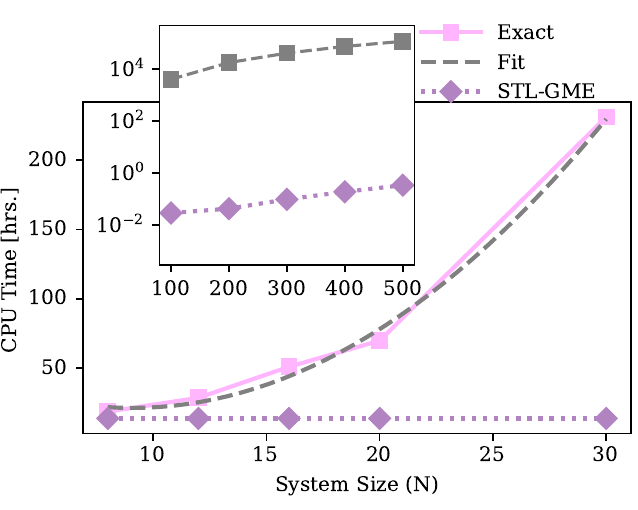}}
    \vspace{-20pt}
\end{center}
\caption{\label{fig:comp-time} Comparison of computational resources (CPU time) needed to reach $25$~ps of simulation time using a direct HEOM simulation versus our STL-GME-enabled simulation. Inset: log scale plot of computational time requirement for the STL-GME extension and (extrapolated fit to) exact simulation for large $N$.}
\vspace{-5pt}
\end{figure}

\textit{\textbf{Computational efficiency:}} The major advantage of our STL-GME is the significant computational savings it offers. Specifically, exact numerical methods scale polynomially or exponentially with $N$ and simulation time. In contrast, our STL-GME scales only as $N^2$ with a prefactor that heavily suppresses the cost to keep it effectively constant (see Fig.~\ref{fig:comp-time}) and (sub)linearly~\footnote{The sublinear scaling in time arises from the ability to exploit the group property of the generator to obtain an effective generator over a larger timestep beyond the onset of Markovian behavior: $\boldsymbol{\mathcal{U}}(n\delta t) = [\boldsymbol{\mathcal{U}}(\delta t)]^n$} in time. The main source of computational expense in our STL-GME arises from the short-time reference calculation on a small lattice needed to construct the generator. Figure~\ref{fig:comp-time} compares the computational cost of HEOM and the STL-GME built from a small reference calculation as a function of $N$. We fit the time costs for HEOM with a polynomial regression (gray dashed line in Fig.~\ref{fig:comp-time}). Even with our highly optimized implementation of HEOM that exploits recent advances in dynamic filtering of auxiliary density matrices~\cite{shi_efficient_2009} and the $n$-particle approximation~\cite{song_time_2015}, the direct simulation for a $100$-site lattice over $25$~fs is $\sim 300$~times more computationally expensive than our STL-GME for the same parameter regime. We can further extend the $100$-site lattice's dynamics to arbitrarily long times using our STL-GME for the trivial cost of repeated small matrix multiplications. This STL-GME-enabled simulation allows us to find that finite-size effects emerge at $\sim 65$ ps for the $100$-site lattice---a timescale that is currently unreachable for a system of this size with a direct HEOM simulation. Such a calculation would require at least $10,500$ CPU hours (optimistically assuming a linear relationship between simulation time and CPU time), making it $750$~times more computationally costly than our STL-GME. Instead, with our STL-GME, the same $100$-site simulation requires effectively the same amount of time as an $820$~fs-long simulation for an $8$-site lattice ($13.5$ CPU hours), as we can compute the STL-GME dynamics in \textit{less than a minute}.   

\textbf{\textit{2D polaron transport}:} Our STL-GME method can easily generalize to higher-dimensional lattices. This enables us to access, for the first time, the exact quantum dynamics of polaron formation and transport in dimensions above one. Accessing transport in 2D is particularly significant as it yields the spatiotemporal spread of polarons on a surface---a phenomenon that recent microscopy experiments measure~\cite{chernikov_exciton_2014, wan_cooperative_2015, yoon_direct_2016, guo_long-range_2017, kennedy_ultrafast_2017, hill_screened_2017, delor_imaging_2020, ginsberg_spatially_2020}. Such simulations can elucidate the microscopic factors that influence energy and charge flow in materials and offer insights on how to modify them for specific applications, such as energy storage or battery development~\cite{bueno_charge_2020, zhang_charge_2023, liu_atomic-scale_2023, schwarz_polaron-based_2024}.

\begin{figure}[b]
\begin{center} 
\vspace{-3pt}
    \resizebox{.49\textwidth}{!}{\includegraphics[trim={5pt 5pt 5pt 5pt},clip]{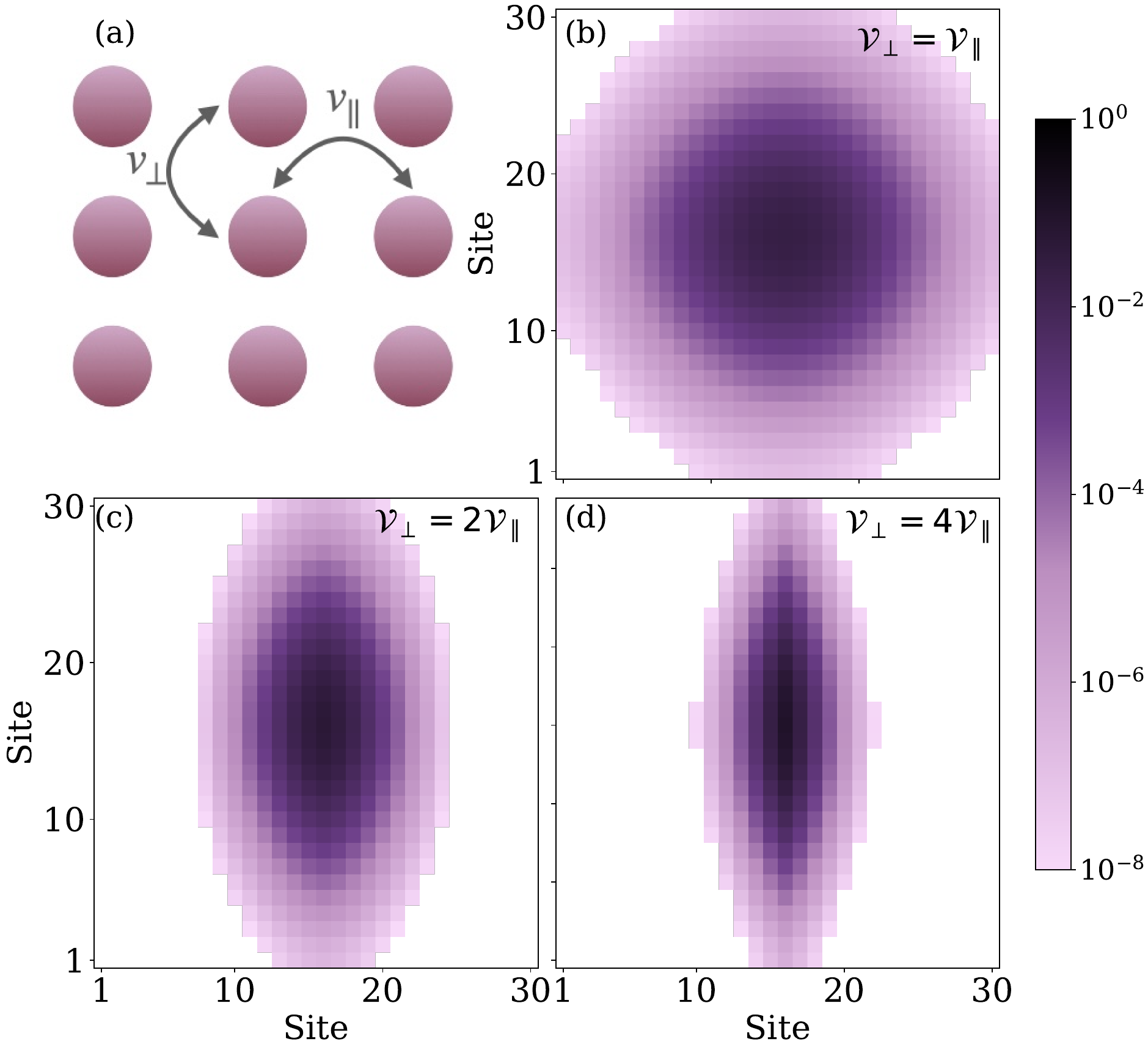}}
    \vspace{-25pt}
\end{center}
\caption{\label{fig:span-2d} Polaron density in a $30 \times 30$-site lattice with dispersive Holstein parameters  $\eta=500$~cm$^{-1}$, $v = 25$~cm$^{-1}$ and  $\omega_c= 41$~cm$^{-1}$. (a) Schematic of a 2D lattice with two types of hopping integrals: $\mathcal{V}_{\perp}$ and $\mathcal{V}_{\parallel}$. Polaron density at $t=10$ ps for three cases: (b) $\mathcal{V}_{\parallel} = \mathcal{V}_{\parallel} = v$; (c) $\mathcal{V}_{\parallel} = v/2$ and $\mathcal{V}_{\perp} = v$; and (d) $\mathcal{V}_{\parallel} = v/4$ and $\mathcal{V}_{\perp} =  v$. }
\vspace{-5pt}
\end{figure}

For our demonstration, we select Hamiltonian parameters characterized by high reorganization energy relative to the hopping integral and a fast phonon decorrelation speed, appropriate for polarons in organic crystals~\cite{troisi_prediction_2007}. To interrogate the effect of various energy scales on polaron transport, we examine a homogeneous 2D lattice characterized by two types of hopping integrals that can cause anisotropic flow, $\mathcal{V}_{\perp}$ and $\mathcal{V}_{\parallel}$, as illustrated in Fig.~\ref{fig:span-2d} (a). Figure~\ref{fig:span-2d} (b)-(d) illustrates the polaron motion on a $30 \times 30$-site 2D lattice with a total of 900 sites. Such a large-scale simulation is made possible exclusively through our STL-GME method (see App.~\ref{benchmark:2d} for implementation details). 

We turn to the effect of varying $\mathcal{V}_{\perp}$ relative to $\mathcal{V}_{\parallel}$. Unsurprisingly, when $\mathcal{V}_{\perp} = \mathcal{V}_{\parallel}$, we observe a symmetric spread of the polaron at $t = 10$ ps, as shown in Fig.~\ref{fig:span-2d} (b). Anisotropic motion can be expected to emerge when $\mathcal{V}_{\perp} \neq \mathcal{V}_{\parallel}$. Indeed, when $\mathcal{V}_{\perp} = 2 \mathcal{V}_{\parallel}$ and $\mathcal{V}_{\perp} = 4 \mathcal{V}_{\parallel}$, polaron density spreads more rapidly along the vertical axis, as shown in Figs.~\ref{fig:span-2d} (c) and (d), respectively, with the ratio $\mathcal{V}_{\perp}/\mathcal{V}_{\parallel}$ dictating the anisotropy of the polaron distribution on the 2D surface. This demonstrates that STL-GME can efficiently simulate nonequilibrium polaron motion in higher-dimensional systems, even in the thermodynamic limit. Hence, this framework offers an accurate and efficient means to simulate polaron transport over the length and time scales needed to compare to experiments. 

\textit{\textbf{Time nonlocal formulation:}} Our space locality arguments are also compatible with the time-nonlocal description of memory~\cite{nakajima_quantum_1958, zwanzig_ensemble_1960, mori_transport_1965}. In App.~\ref{time-nonlocal}, we formulate the space-local time-nonlocal (SL-TNL) GME, where we further unify the SL framework with the Transfer Tensor Method~\cite{cerrillo_non-markovian_2014}. Interestingly, we find that the time-local characteristic memory distance, $\mathrm{d}_{\mc{U}}$, is shorter than its time-nonlocal counterpart, $\mathrm{d}_{\mc{K}}$. We also find that the generator and memory kernel lifetimes in the time-local and time-nonlocal formulations follow a similar inequality, $\tau_{\mc{U}} \leq \tau_{\mc{K}}$, consistent with our previous findings in biomolecular dynamics~\cite{dominic_building_2023}. 

\begin{figure}[t]
\begin{center} 
\vspace{-3pt}
    \resizebox{.49\textwidth}{!}{\includegraphics[trim={0pt 0pt 0pt 0pt},clip]{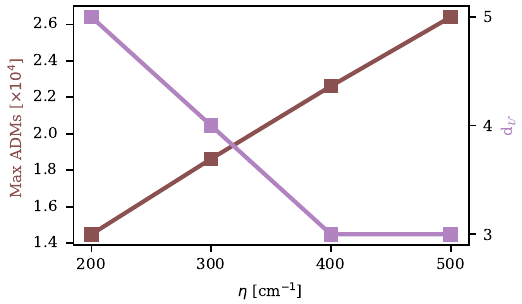}}
    \vspace{-25pt}
\end{center}
\caption{\label{fig:adm} The computational cost of HEOM (solid brown line) rises with growing electronic-phonon coupling ($\eta$) while the cutoff (and therefore cost) decreases for our STL-GME (solid purple line). Hamiltonian parameters: $v = 50$ cm$^{-1}$, $\omega_c = 100$ cm$^{-1}$.}
\vspace{-15pt}
\end{figure}

\textit{\textbf{Entanglement:}} For systems involving coupling between spins or fermions with phonons, increasing the reorganization energy leads to high electronic-nuclear entanglement. This generally translates into greater computational cost for many numerically exact~\cite{tanimura_time_1989, makri_tensor_1995, wang_systematic_2001,thoss_self-consistent_2001, wang_multilayer_2015} and approximate dynamics methods~\cite{stock_semiclassical_1997, sun_semiclassical_1998, miller_electronically_2009}. For instance, the number of differential equations that must be propagated at each step in HEOM grows significantly with increasing reorganization energy, raising its computational cost. Similarly, in matrix product state-based propagation schemes, the bond dimension increases with increasing reorganization energy and simulation time~\cite{kloss_multiset_2019, fux_tensor_2023}, further escalating computational demands. In contrast, in our SL-GME approach, increasing reorganization energy \textit{reduces} $\mathrm{d}_{\mathcal{U}}$ (see Fig.~\ref{fig:adm}). This allows us to perform reference calculations on smaller lattices, providing dramatic computational savings in systems with large reorganization energies.

\section{Conclusion}
\vspace{-4pt}
While it is well-known that the temporal truncation of memory in traditional GMEs can significantly lower the computational cost of dynamic simulations \textit{over arbitrary times}, here we introduce space-local GMEs where spatial truncation of memory can dramatically lower the cost of accessing the dynamics of complex many-body systems \textit{of arbitrary sizes}. By integrating both spatial and temporal truncation in the same GME framework within a time-local framework, we have developed a novel STL-GME that allows us to leverage the dynamics of small lattices over short times to exactly simulate the dynamics of thermodynamically large systems across arbitrary times, free from finite-size effects. In addition to our time-local formulation, we have developed analogous continuous and discrete time-nonlocal GMEs. 

We have demonstrated the benefits offered by our STL-GME when simulating polaron formation and transport in dispersive Holstein lattices in 1D and 2D with short-range couplings. In fact, our STL-GME has already revealed that nonequilibrium lattice relaxation can exponentially delay the onset of polaron diffusion~\cite{bhattacharyya_nonequlibrium_2024}. Our approach, however, can be expected to find applications across a wide range of dissipative lattice problems to interrogate the nonequilibrium relaxation dynamics of charge, electronic energy, and heat flow. We have shown, for example, that our SL-GMEs are broadly compatible with systems displaying short-range interactions and in arbitrary dimensions. In addition, like previous GMEs, our SL-GMEs are agnostic to the choice of dynamics solver (classical or quantum, exact or approximate) used to construct the time-local generator or memory kernel, broadening its applicability to a wide variety of systems and phenomena. By offering access to experimentally relevant nonequilibrium relaxation processes in thermodynamically large systems over arbitrarily long times, our STL-GME can be expected to provide a transformative tool to uncover and explain new dynamical phenomena in periodic materials.

\section*{Acknowledgements}
\vspace{-4pt}
Acknowledgment is made to the donors of the American Chemical Society Petroleum Research Fund for partial support of this research (No.~PRF 66836-DNI6). A.M.C.~acknowledges the support from a David and Lucile Packard Fellowship for Science and Engineering. S.B.~acknowledges the John Bailar Memorial Endowment and the Marion L. Sharrah Fellowship for partial support of the research. T.S.~is the recipient of an Early Career Fellowship from the Leverhulme Trust. We thank Prof.~Qiang Shi for sharing his HEOM code with us. This work utilized the Alpine high-performance computing resource at the University of Colorado Boulder. Alpine is jointly funded by the University of Colorado Boulder, the University of Colorado Anschutz, Colorado State University, and the National Science Foundation (award 2201538).

\section*{Data Accessibility}
\vspace{-4pt}
The data supporting our study's findings are available from the corresponding author upon reasonable request.

\appendix

\section{HEOM simulations}
\label{HEOM-details}

HEOM~\cite{tanimura_time_1989} is a numerically exact method that predicts the dynamics of the reduced density matrix of an open quantum system by mapping environmental degrees of freedom to auxiliary density matrices (ADMs) that quantify the number of coupled differential equations being solved simultaneously.  The reduced density matrix provides information about the polaron population.

We converge all HEOM calculations with respect to the hierarchical depth $L$, the number of Matsubara frequencies $K$, and the time step $\delta t$. For our 1D simulations, $L = 22$, $K = 1$, and $\delta t = 0.25$ fs. When changing reorganization energy for Fig.~\ref{fig:adm}, we use $L = 26$, $K = 2$, and $\delta t = 0.25$ fs. Due to the extreme computational cost, we employed a high-temperature approximation for 2D simulations, setting $K = 0$, and converged the hierarchical depth and timestep to $L = 26$ and  $\delta t = 0.25$ fs. In all our simulations, we apply dynamic filtering~\cite{shi_efficient_2009}, using a threshold of $N_{\text{cut}} = 10^{-7}$ atomic units, in line with previous work~\cite{yan_theoretical_2019, bhattacharyya_anomalous_2024}. We also employ $n$-particle approximation~\cite{song_time_2015} to minimize numerical costs without affecting the accuracy of the results.

\section{GME Formulation} \label{app:pop-GME}

At the heart of GMEs is the projection operator~\cite{grabert_projection_1982}, $\mathcal{P}$, which enables one to derive the GME and establish the form of the correlation function of interest. Because we are interested in tracking the population matrix, we choose the nonequilibrium population projector~\cite{sparpaglione_dielectric_1988, montoya-castillo_approximate_2016} $\mathcal{P} = |\boldsymbol{A})(\boldsymbol{A}| = \sum_{j=1}^N |A_j) (A_j|$, where $A_j = \hat{a}_j^\dag \hat{a}_j$ is the operator tracking fermionic occupation of site $j$. As in previous work~\cite{montoya-castillo_approximate_2016}, we define the inner product as 
\begin{equation}
    (A_i|\mathcal{\hat{O}} |A_j) = \frac{1}{Z_{\rm ph}} \mathrm{Tr} \Big[ A_i^\dag \mathcal{\hat{O}} A_j e^{-\beta \hat{H}_{ph}}\Big],
\end{equation}
where $\mathcal{\hat{O}}$ is a superoperator and $Z_{\rm ph}$ is the partition function of the phonon modes. With this choice of projection operator, we can obtain the population correlation matrix $\boldsymbol{C} (t)$ as follows 
\begin{equation}
\begin{split}
  \boldsymbol{C} (t) & = (\boldsymbol{A}| e^{-i\mathcal{L} t}|\boldsymbol{A}),
\end{split}
\end{equation}
where $\boldsymbol{C} (0) = \mathbb{1}$. Here, $e^{-i\mathcal{L}t}$ is the propagator that evolves an initial density in time, and $\mathcal{L}= [H, ...]$ is the Liouvillian operator. The $i,j$ element of this correlation function takes the form
\begin{equation}
\begin{split}
    C_{i,j}(t)&= \frac{1}{Z_{\rm ph}} \mathrm{Tr} \Big[ A_i^\dag e^{-\mathrm{i}\mathcal{L}t} A_j e^{-\beta \hat{H}_{ph}}\Big] \\
    & =  \mathrm{Tr} \Big[\hat{a}_i^\dag \hat{a}_i e^{-\rmm{i}\mathcal{L}t} \hat{\rho}_j(0) \Big].
\end{split}
\end{equation}
We calculate $\boldsymbol{C} (t)$ using HEOM.

\section{STL-GME for a 1D lattice}\label{app:stl-gme}
\vspace{-5pt}
Here, we outline our protocol for building a size-augmented STL-GME for a 1D lattice using only a short-time simulation of a small lattice. Here we use HEOM to generate the reference dynamics, but our method is agnostic to the choice of solver.
\begin{enumerate}
    \item Construct the generator $\boldsymbol{\mathcal{U}}(t)$ from the reference $\boldsymbol{C}(t)$ using Eq.~\ref{eq-generator}.

    \item Identify the lifetime $\tau_{\mc{U}}$ by computing the $||\mathrm{L}||_2$ error between reference and GME-generated dynamics as a function of the proposed lifetime, $\tau$, $\mathrm{Error}(\tau) =  \frac{1}{N_t} ||\boldsymbol{C}^{\rm ref}(t) - \boldsymbol{C}^{\rm GME}(t; \tau) ||_2 $, and identify when the error function plateaus to a sufficiently low value. $ N_t $ represents the number of time points predicted by TL-GME, i.e., the number of time points used to construct the generator subtracted from the total number of time points in the reference dynamics. Our threshold per element of $\boldsymbol{C} (t) $ is $3 \times 10^{-8}$.

    \item Identify characteristic memory distance $\mathrm{d}_{\mc{U}}$ in a similar way by analyzing the error between reference dynamics and GME dynamics via the error as a function of the proposed cutoff distance, $d$, $\mathrm{Error}(d) = \frac{1}{N_t} ||\boldsymbol{C}^{\rm ref}(t) - \boldsymbol{C}^{\rm GME}(t; d) ||_2 $. For SL-GME, $N_t$ is the total number of points. Our threshold per element of $\boldsymbol{C} (t) $ is $6 \times 10^{-8}$.

    \item If one finds $\mathrm{d}_{\mc{U}} < N/2$, this means reference simulation is sufficiently large for system size extension via generator augmentation. Keep the generator up to its lifetime $\tau_{\mc{U}}$ only. Truncate the generator by setting its entries connecting sites with a relative distance $d > \mathrm{d}_{\mc{U}}$ to 0.

    \item Augment the generator's dimension from $[N, N, tsteps]$ to $[M, M, tsteps]$, where $M$ is the length of the extended lattice. Then, populate all additional new entries in the matrix with $0$.

    \item Employ $\boldsymbol{\mathcal{U}}(t) [M, M] $ to propagate the dynamics. For time $t\leq \tau_{\mc{U}}$, use Eq.~\ref{TL-GME}. For $t> \tau_{\mc{U}}$, use Eq.~\ref{TL-GME2}.
\end{enumerate}

\begin{figure}[t]
\begin{center} 
\vspace{-3pt}
    \resizebox{.49\textwidth}{!}{\includegraphics[trim={0pt 0pt 0pt 0pt},clip]{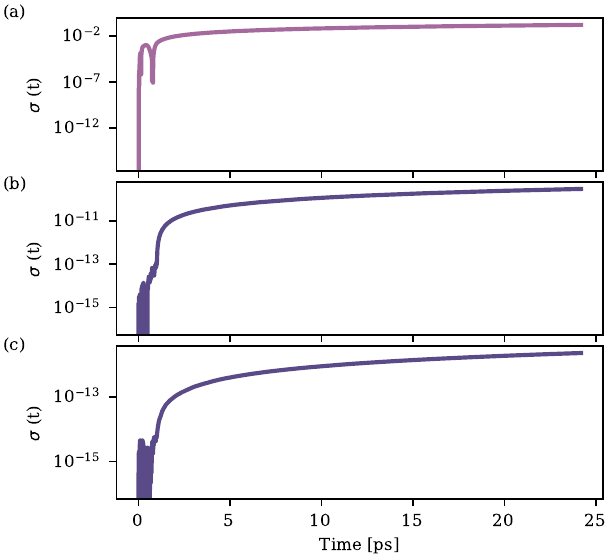}}
    \vspace{-20pt}
\end{center}
\caption{\label{fig:consv} Population loss $\sigma(t)$ due to spatial truncation of the generator elements for the STL-GME dynamics for a 20-site lattice built from a reference 8-site simulation, as shown in Fig.~\ref{fig:space-time-local}. Population loss as a function of simulation time arising from: (a) performing a direct truncation of the generator; (b) adopting the redistribution scheme; (c) adopting the renormalization scheme}
\vspace{-10pt}
\end{figure}

\textbf{\textit{Ensuring population conservation:}} In our STL-GME method, we discard the elements of $\boldsymbol{\mc{U}}$ with relative lattice spacings larger than $d_{\mc{U}}$ by setting them to $0$. Our generator truncation---like singular value truncations in tensor network-based methods~\cite{greene_tensor-train_2017}---can cause violations of population conservation. We track this population loss via $\sigma (t) = |1 - \sum_k C_k(t)|$. Ideally, $\sigma(t)$ should be as close to zero as possible. Figure~\ref{fig:consv} (a)---which corresponds to the parameter regime of Fig.~\ref{fig:space-time-local} where we employ an 8-site reference simulation to predict the dynamics of a 20-site lattice---reveals that our spatial truncation causes a $~1\%$ population loss over the first 1~ps, which will continue to grow with simulation time.

To conserve the population, we propose two schemes:
\begin{enumerate}
    \item \textit{Redistribution:} Add all discarded elements and distribute them equally among all remaining elements of $\boldsymbol{\mathcal{U}}$ after truncation,
    \begin{align}\label{eq:redistribure}
    \mathcal{U}[i,j; d < \mathrm{d}_{\mathcal{U}}] = & \: \mathcal{U}[i,j; d < \mathrm{d}_{\mathcal{U}}] \nonumber \\
    & + \frac{1}{N_{d_{\mathcal{U}}}} \sum_j \mathcal{U}[i,j; d > \mathrm{d}_{\mathcal{U}}], \quad \forall i,
    \end{align}
    where $ N_{d_{\mathcal{U}}} $ is the number of elements in $ \mathcal{U} $ that we keep as nonzero entries in our generator after spatial truncation.

    \item \textit{Renormalization:} After spatial truncation, renormalize the generator $\boldsymbol{\mathcal{U}} $ matrix at each time step,
    \begin{equation}\label{eq:renormalize}
    \mathcal{U}[i,j; d < \mathrm{d}_{\mathcal{U}}] = \frac{\mathcal{U}[i,j; d < \mathrm{d}_{\mathcal{U}}]}{\sum_j \mathcal{U}[i,j; d < \mathrm{d}_{\mathcal{U}}]}, \quad \forall i.
    \end{equation}
\end{enumerate}

Figure~\ref{fig:consv} shows when we add our population conservation schemes by redistributing or renormalizing spatially truncating generator $\boldsymbol{\mathcal{U}}$ we can keep the population loss in the order of $10^{-11}$ and $10^{-12}$, even for longer simulation time. Figure~\ref{fig:consv} (b) and (c) shows the population loss quantity for redistribution and renormalization scheme respectively. In our study, we employ redistribution scheme for all 1D simulations and renormalization scheme for 2D simulations. While we show that we modify the generator to conserve the population, one could, in principle, imagine some level of modification in the population level to conserve the population.

\begin{figure}[t]
\begin{center} 
\vspace{-3pt}
    \resizebox{.49\textwidth}{!}{\includegraphics[trim={0pt 0pt 0pt 0pt},clip]{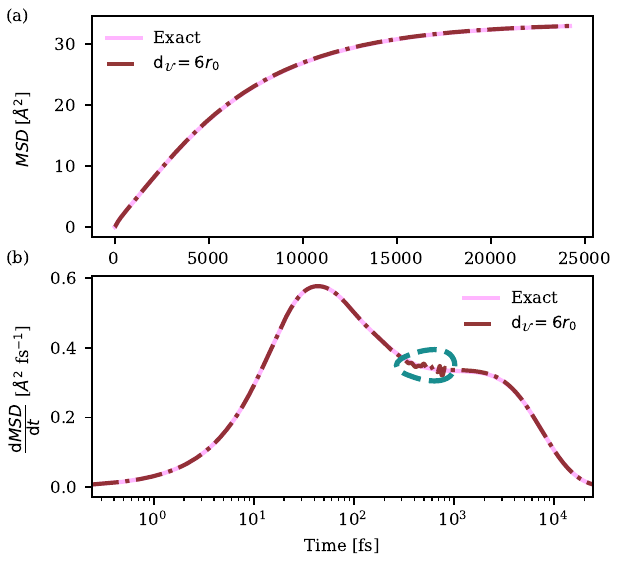}}
    \vspace{-20pt}
\end{center}
\caption{\label{fig:msd_d6} (a) Comparison of $MSD$ between exact result and STL-GME dynamics data for Fig.~\ref{fig:space-local} parameters. Here we set a characteristic memory distance $\mathrm{d}_{\mc{U}} = 6 r_0$. (b) Comparison of $\mathrm{d}MSD/\mathrm{d}t$ between the exact result and the STL-GME dynamics with $\mathrm{d}_{\mc{U}} = 6 r_0$ for Fig.~\ref{fig:space-local} parameters. The green encircled region between 400 fs and 1000 fs shows the inclusion of noise in the STL-GME dynamics. }
\vspace{-10pt}
\end{figure}

\textbf{\textit{Managing finite numerical precision of the dynamical solver}:} Figure~\ref{fig:schematic} (a) shows that increasing lattice distance $d$ makes the elements of the generator $\boldsymbol{\mc{U}}$ smaller. At some point, the size of these generator elements is commensurate with the statistical error or numerical precision error of the reference dynamical solver. While keeping the full generator intact ensures that the generator recovers the reference dynamics, truncating generator elements that may be at the level of the solver error can introduce small disagreements between the STL-GME and reference dynamics because of imperfect cancellation of error. 

\begin{figure}[b]
\begin{center} 
\vspace{-3pt}
    \resizebox{.49\textwidth}{!}{\includegraphics[trim={0pt 0pt 0pt 0pt},clip]{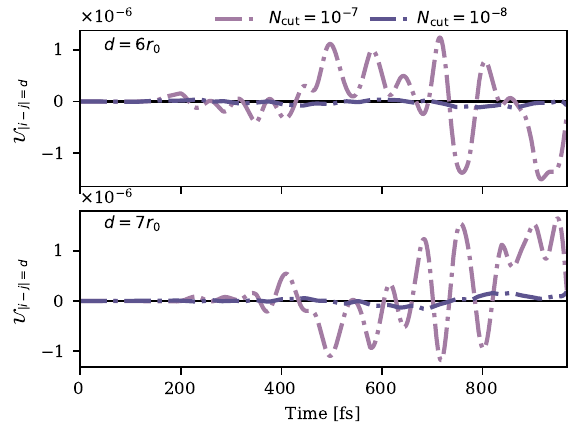}}
    \vspace{-20pt}
\end{center}
\caption{\label{fig:u_d7} Generator elements of $\mc{U}$ for inter-site distances $d=6r_0$ and $d=7r_0$ for the parameters in Fig.~\ref{fig:time-local} as a function of the strength of dynamic filtering in the HEOM solver, $N_{\rm cut}$.}
\vspace{-10pt}
\end{figure}

To see how this finite precision error in the underlying solver can manifest in the STL-GME, consider generator elements connecting sites separated by inter-site distances of $6r_0$ and $7r_0$ in Fig.~\ref{fig:u_d7} (purple lines), which have contributions on the order of $10^{-6}$ to $10^{-7}$. In our reference HEOM simulations, from which we build the generator, we use dynamic filtering with a threshold of $N_{\rm cut} = 10^{-7}$ atomic units. Including these small noisy elements in the truncated $\boldsymbol{\mc{U}}$ risks introducing noise into our dynamical quantities, with derivative quantities magnifying the effect of such noise. Specifically, when we choose $d_{\mc{U}} = 6 r_0$ and compare the exact and STL-GME-predicted $MSD$, as shown in Fig.~\ref{fig:msd_d6} (a), one cannot notice any visual difference. In contrast, when we compare the more sensitive $dMSD/dt$, we find a minor noisy behavior in the green-encircled region of Fig.~\ref{fig:msd_d6} (b). This noise arises because elements in $\mathcal{U}$ for $d \geq 6 r_0$ are discarded and set to $0$, leading to insufficient noise cancellation from elements with $d = 7 r_0$ and beyond. Reducing this filter threshold to $N_\text{cut}$) to $10^{-8}$ decreases the noise in the $6 r_0$ and $7 r_0$ elements of $\boldsymbol{\mathcal{U}}$ (see Fig.~\ref{fig:u_d7}). Hence, if smaller elements need to be included in the generator when implementing STL-GME, one must be mindful of the numerical precision of the underlying dynamical solver.

\section{STL-GME for a 2D lattice}\label{benchmark:2d}
\vspace{-5pt}

Here we summarize our modified protocol to build the STL-GME for a homogeneous 2D lattice. We note, however, that the same protocol can be used for an $N$D lattice with $N \geq 2$. To construct the input dynamical matrix, $\boldsymbol{C}(t)$, we perform a single reference HEOM simulation and then exploit translational symmetry. Once we construct $\boldsymbol{C}(t)$, we follow the following workflow to implement the STL-GME for a 2D lattice:

\begin{enumerate}
    \item Compute the generator $\boldsymbol{\mc{U}} (t)$ from reference $\boldsymbol{C} (t)$ using Eq.~\ref{eq-generator}.
    \item Identify the generator lifetime $\tau_{\mc{U}}$ by comparing the $||\mathrm{L}||_2$ error between TL-GME results and reference dynamics.
    \item Identify the spatial distance cutoff $\mathrm{d}_{\mc{U}}$:
    \begin{enumerate}
        \item Keep the generator up to its lifetime, ${\mc{U}} [N, N, t \leq \tau_{\mc{U}}]$.
        \item Reshape the first two lattice indices into $x$ and $y$ coordinate indices such that any lattice index $a$ can be rewritten as 
        \begin{equation}\label{eq:index-collapse}
        a = i\times N_x + j,
        \end{equation}
        where $N = N_x \times N_y$. This transforms the generator into a 5-rank tensor: ${\mc{U}} [N, N, t] \to {\mc{U}} [N_x, N_y; N_x, N_y, t]$.
        \item To spatially truncate the generator for a homogeneous lattice: 
        \begin{enumerate}
            \item Select an excitation starting at any lattice point, e.g., the origin of the 2D lattice, ${\mc{U}} [N_x, N_y; 0, 0, t]$.

            \item Compute the distance between this initial excitation at coordinates $(0,0)$ and any measurement site $(i,j)$ as
        \begin{equation}\label{eq:def-distance}
        d_{i,j} = \sqrt{i^2 + j^2} \quad \forall i,j.
        \end{equation}
        \item Discard all the elements where $d_{i,j} > \mathrm{d}_{\mc{U}}$ by making them numerically 0.
        \item Apply Eq.~\ref{eq:renormalize} to this spatially truncated ${\mc{U}} [N_x, N_y; 0, 0, t]$ to conserve the population.
        \item Employ translational symmetry to populate full ${\mc{U}} [N_x, N_y; N_x, N_y, t]$ from ${\mc{U}} [N_x, N_y; 0, 0, t]$ by varying initial excitation lattice coordinates.
        \end{enumerate}
        \item Collapse the coordinate indices into lattice indices using Eq.~\ref{eq:index-collapse} such that ${\mc{U}} [N_x, N_y; N_x, N_y, t] \to {\mc{U}} [N, N, t].$
        \item Use the modified generator to propagate the dynamics using Eq.~\ref{TL-GME} (before $ t < \tau_{\mc{U}}$) and Eq.~\ref{TL-GME2} after $ t < \tau_{\mc{U}}$.
        \item Calculate the $||\mathrm{L}||_2$ error metric by comparing the exact dynamics with GME-predicted dynamics for different potential distance cutoffs to find $\mathrm{d}_{\mc{U}}$.
    \end{enumerate}
    \item If $\mathrm{d}_{\mc{U}} < N/2$, the reference calculation is sufficiently large for spatial extension of the dynamics. To generate the dynamics of a system with extended size:
    \begin{enumerate}
        \item Augment the generator's dimension. Expand generator obtained from step (3c) ${\mc{U}} [N_x, N_y; 0, 0, t]$ to ${\mc{U}} [M_x, M_y; 0, 0, t]$ for all time points, where $M_x \ge N_x$, $M_y \ge N_y$ and $M = M_x \times M_y.$, and assign $0$ to all the new entries.
        \item Similar to step (3d), use translational symmetry to populate ${\mc{U}} [M_x, M_y; M_x, M_y, t]$ for all initial conditions from ${\mc{U}} [M_x, M_y; 0, 0, t]$ and collapse the coordinate indices to lattice indices, yielding  the expanded generator, ${\mc{U}} [M, M, t]$, for the $M \times M$ lattice system.
        \item Similar to step (4d), propagate the GME dynamics using ${\mc{U}} [M, M, t]$ which predict population correlation matrix $\boldsymbol{C} (t) $ for $M \times M$ lattice.
    \end{enumerate}
\end{enumerate}
For Figs.~\ref{fig:span-2d} (b)-(d), we simulate a $8 \times 8$ lattice to obtain the reference $\boldsymbol{C}(t)$ dynamics for the STL-GME extension. We find that the generator lifetime $\tau_{\mc{U}} = 800$~fs and appropriate distance cutoff $\mathrm{d}_{\mc{U}} = 3 r_0$, suggesting that the $8 \times 8 $ lattice is sufficient as the reference simulation. We employ this $8 \times 8 $ dynamics to predict the dynamics of a $10 \times 10$ lattice and confirm that our STL-GME dynamics agree with a separate exact HEOM simulation. In the main text (see Fig.~\ref{fig:span-2d} (b)-(d)), we employ our STL-GME protocol to augment the size to a $ 30 \times 30 $ lattice and simulate over the first $ 100 $~ps.

\begin{figure}[t]
\begin{center} 
\vspace{0pt}
    \resizebox{.49\textwidth}{!}{\includegraphics[trim={0pt 0pt 0pt 0pt},clip]{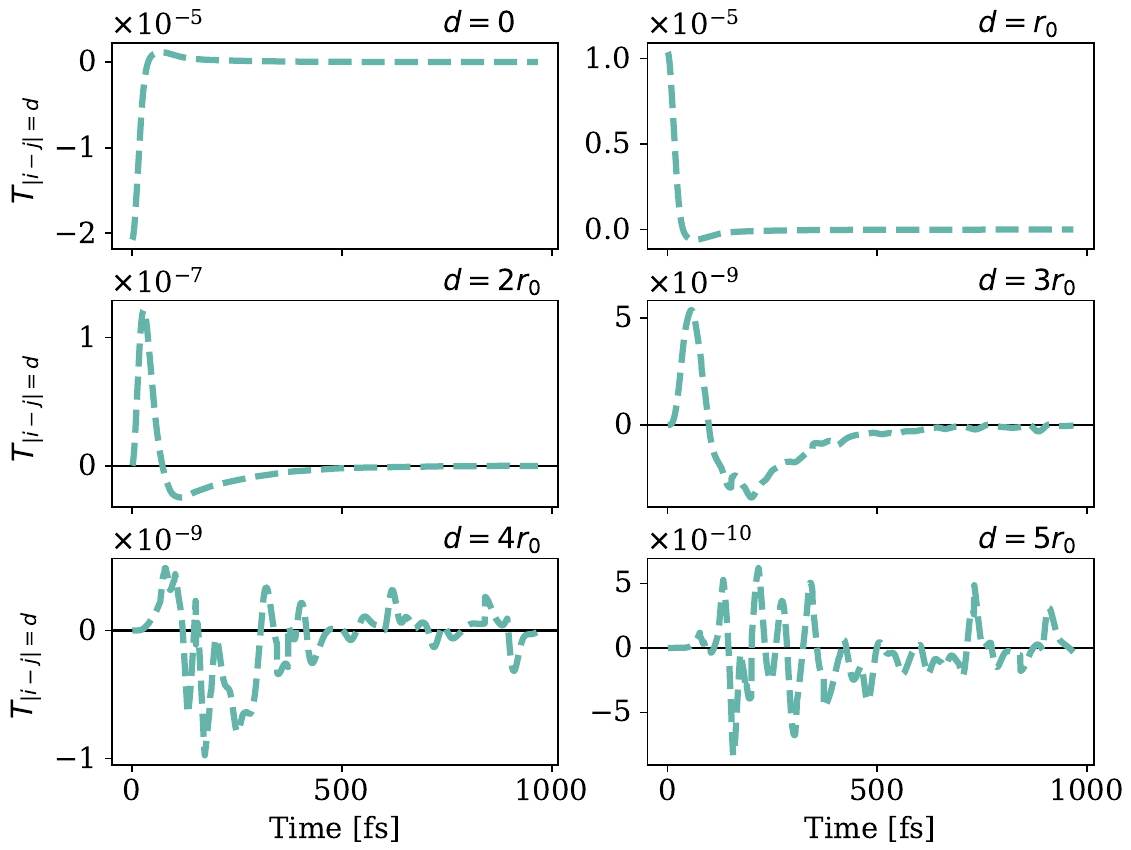}}
    \vspace{-20pt}
\end{center}
\caption{\label{fig:t-tensor}  Elements of the transfer tensor, $\boldsymbol{T}(t) $, as a function of intersite distance $d$ for the dispersive Holstein lattice with the same parameters as Fig.~\ref{fig:time-local}.}
\end{figure}

\section{Space-local time-nonlocal GME}
\label{time-nonlocal}
\vspace{-5pt}
Here we discuss the spatial locality of the memory kernel in the time-local GME and how to truncate small elements that become negligible with increasing inter-site distance. For lattice problems with short-range interactions and local couplings (such as the dispersive Holstein lattice), space-locality is also present in the time-nonlocal description of memory. We demonstrate space-locality in the Transfer Tensor Method (TTM), which is the discrete analog of the time-nonlocal GME~\cite{cerrillo_non-markovian_2014}.

\begin{figure}[t]
\begin{center} 
\vspace{-3pt}
    \resizebox{.49\textwidth}{!}{\includegraphics[trim={0pt 0pt 0pt 0pt},clip]{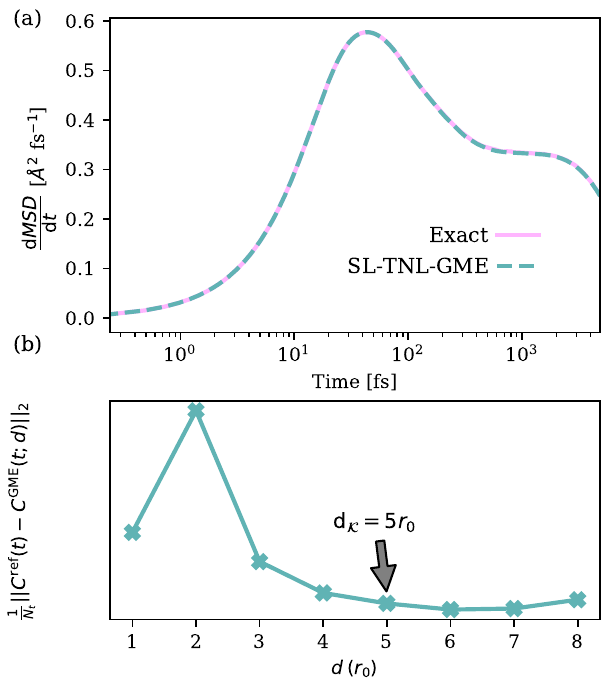}}
    \vspace{-20pt}
\end{center}
\caption{\label{fig:sl-tnl} Performance of the SL-TNL-GME for a dispersive Holstein lattice with $N=20$ sites and $\eta=323$~cm$^{-1}$, $v = 50$~cm$^{-1}$ and $\omega_c= 41$~cm$^{-1}$. (a) Our SL-TNL-GME (green dashed lines) recapitulates the reference simulation of $\mathrm{d}MSD/\mathrm{d}t$ starting at $d_{\mc{K}} = 5 r_0$. (b) Identification of the characteristic memory distance in the time-nonlocal framework from the deviation of the SL-TNL-GME from the reference dynamics as a function of the proposed distance cutoff. }
\vspace{-10pt}
\end{figure}

For the population projector, one can write the integrated time-nonlocal master equation via the TTM formulation, 
\begin{equation}\label{eq:ttm}
    \boldsymbol{C}(t) = \sum_{k} \boldsymbol{T}(t-k) \boldsymbol{C}(k),
\end{equation}
where the memory kernel of the dynamics is encoded into the transfer tensor, $\boldsymbol{T}(t)$. We observe that the elements of the $T_{i,j}(t)$ tensor become smaller with increasing lattice distance $d$, (see Fig.~\ref{fig:t-tensor}). Hence, we can spatially truncate $\boldsymbol{T} (t)$ and then augment it to a larger dimension by adding zeroes to the new entries.

To demonstrate the performance of the SL-TNL-GME, we focus on the parameter regime of Fig.~\ref{fig:time-local}. Figure~\ref{fig:sl-tnl} (a) confirms that the SL-TNL-GME can recover the reference simulation. Figure~\ref{fig:sl-tnl} (b) shows that the characteristic memory distance in the time-nonlocal framework is $\mathrm{d}_{\mathcal{K}} = 5 r_0$, which is bigger than the analogous characteristic memory distance in the time-local framework, $\mathrm{d}_{\mathcal{U}} = 3 r_0$. Similarly, we find the memory kernel lifetime in the time-nonlocal formulation is $\tau_{\mathcal{K}} = 850$~fs, which is bigger than the lifetime we identify in the time-local case, $\tau_{\mathcal{U}} = 820$~fs. The latter result is consistent with previous findings in the context of biomolecular dynamics~\cite{dominic_building_2023}. Hence, this example shows one can implement the idea of spatial memory truncation in the time-nonlocal formulation of reduced dynamics.
\vfill
\pagebreak

\section*{References}

\vspace{-14pt}
\bibliography{references}

\end{document}